\documentclass[preprint2]{aastex}
\usepackage{url}
\usepackage{natbib}
\usepackage{graphicx}
\usepackage{color,soul}
\usepackage[english]{babel}

\shorttitle{VisIVOWeb : http://visivoweb.oact.inaf.it}
\shortauthors{Costa A. et al.}
\begin{document}
\title{VisIVOWeb: A WWW Environment for Large-Scale Astrophysical Visualization}
\author{A. Costa \altaffilmark{1}, U. Becciani \altaffilmark{1}, P. Massimino \altaffilmark{1}, M. Krokos \altaffilmark{3},\\ G. Caniglia \altaffilmark{1}, C. Gheller \altaffilmark{2}, A. Grillo \altaffilmark{1, 4} and F. Vitello \altaffilmark{1}} 
\affil{INAF-Osservatorio Astrofisico di Catania, Italy}
\affil{CINECA, Italy}
\affil{University of Portsmouth, United Kingdom}
\affil{Cometa Consortium, Italy}
\email{alessandro.costa@oact.inaf.it}
\begin{abstract}
This paper presents a newly developed WWW environment called VisIVOWeb aiming at providing the astrophysical community with powerful visualization tools for large-scale datasets in the context of Web 2.0. VisIVOWeb can handle effectively modern numerical simulations and real-world observations. Our software is open-source based on established visualization toolkits and offers high quality rendering algorithms. The underlying data management is discussed together with the supported visualization interfaces and movie making functionality. We introduce VisIVOWeb Network, a robust network of customized web portals for visual discovery and VisIVOWeb Connect, a lightweight and efficient solution  for seamlessly connecting to existing astrophysical archives. A significant effort has been devoted for ensuring interoperability with existing tools by adhering to IVOA standards. We conclude with a summary of our work and a discussion on future developments.
\end{abstract}
\section{Introduction}
The primary goal of scientific visualization is to create images and animations to aid scientists in understanding complex datasets, e.g. large-scale multi-dimensional datasets obtained either from numerical simulations  or real-world astrophysical observations.
\\The typical sizes of modern astrophysical datasets are increasing steadily due to continuous advances in computational power and instrumentation. As a result such datasets often require storage in a distributed way, e.g. the Millenium-II simulation \cite{Millennium} which produced nearly 20 TBs. Present and next-generation sky surveys are also of constantly increasing sizes and complexity, e.g. the Sloan Digital Sky Survey ~\cite{SLOAN} that offers a catalogue of 15.7 TBs. Moreover a significant effort has been devoted in handling highly-complex numerical simulations in the context of  the Virtual Observatory ~\cite{VO,NVO}, e.g.  the Italian Theoretical Virtual Observatory\footnote{ \url{http://itvo.oact.inaf.it/ }} ~\cite{ITVO} and the German Astrophysical Virtual Observatory ~\cite{GAVO} offering dark matter halo catalogues and merger trees from the Millennium-II.
\\Modern visualization can aid astrophysicists in gaining good insights of highly-complex datasets, through rapid and intuitive discovery of familiar patterns and correlations between properties, without involving CPU intensive analysis codes. As a result, suitably-constructed visualization tools can be instrumental for future astrophysical advances, e.g. by allowing comparing simulations meaningfully or even correlating appropriately simulations and observational datasets. As an example consider the massive halo formation process, or interaction among galaxies as seen in the evolution of a simulation and similar observed objects\footnote{\url{http://www.mpa-garching.mpg.de/galform/data_vis/ }}.
\\This article presents a WWW environment for visualization of large-scale astrophysical datasets that can be installed easily in data centers hosting the datasets, offering astrophysicists an array of advanced visualization tools exploited effectively using common Internet browsers. VisIVOWeb is our open-source visualization software; it is hosted by sourceforge.net\footnote{ \url{http://sourceforge.net/projects/visivoweb/ }} and is released under GNU\footnote{\url{http://www.gnu.org/licenses/gpl.html}} license. The latest version can be accessed through Subversion\footnote{\url{http://subversion.apache.org/ }}, an open-source revision control system.
\\The related work is outlined in section 2. The main ingredients of VisIVOWeb are then introduced, namely handling of large-scale datasets, data management services and supported visualization and exploration interfaces, and finally VisIVOWeb Network and VisIVOWeb Connect for networking VisIVOWeb portals and connecting them seamlessly to astrophysical archives. The way users can gain access to the system and import their datasets is described in sections 3 and 4 respectively.  Section 5 describes the underlying data-model required to support VisIVOWeb functionality and outlines a typical operational scenario. The visualization algorithms supported together with the relevant visualization interfaces are presented in sections 6 and 7. We discuss the creation of scientific movies in section 8 focusing on a specific example for demonstration. We then outline VisIVOWeb Network and VisIVOWeb Connect together with our recent dissemination activity. Finally we conclude with a summary of our work including pointers to some future developments.

\section{Related works}
VisIVOWeb implements a customized WWW interface wrapped around VisIVOServer that can be used through common Internet browsers. We have implemented VisIVOServer previously  ~\cite{VisIVOServer} through a comprehensive collection of modules for processing and visualization of astrophysical datasets. The software is open-source, written in C++ and hosted by sourceforge.net\footnote{\url{http://sourceforge.net/projects/visivoserver/ }}. The underlying visualization engine is founded on the functionality of the Visualisation Toolkit\footnote{\url{http://www.vtk.org/ }}~\cite{VTK4thEdition}. As VisIVOWeb is wrapped around VisIVOServer, it offers fast rendering of customized 3D views of large-scale astrophysical datasets in a WWW environment that provides a unique mix of data management and multi-dimensional visualization tools.\\
To the best of our knowledge, no other comparable environments exist in the Virtual Observatory ~\cite{VO,NVO}. Typically the rendering process is computationally intensive often requiring millions  of floating-point and integer operations for generating single 3D views ~\cite{An introduction to parallel rendering}. This is the reason for developing a fully C++ implementation rather than optimizing other existing Java-based environments, e.g. TOPCAT ~\cite{TOPCAT}. Compared to TOPCAT our rendering  capability is significantly enhanced as we can handle large-scale mutli-dimensional datasets. The visualization functionality provided by VisIVOWeb is also very different compared to a plotting tool such as VoPlot ~\cite{VOPlot} and Aladin ~\cite{Aladin}. The latter is simply a software sky atlas allowing the user to visualize digitized astronomical images and to also superimpose entries from astronomical catalogues or databases.\\
Among the visualization algorithms provided by VisIVOWeb for handling point-like datasets Splotch ~\cite{Splotch1, Splotch} deserves particular attention. A customized ray-tracer is employed for creating meaningful 3D views of modern cosmological simulations effectively. The intensive underlying calculation is optimized by pre-procesing the relevant particles, and recent developments exploit GPUs for hardware acceleration. The rendering is achieved through composition of the final color in each pixel by calculating emission and absorption of individual volume elements. Splotch is released under GNU license and is integrated in VisIVOServer fully. We focus on visualization interfaces within VisIVOWeb in section 7. The next section outlines the typical user scenario for gaining access to our system.
 
\section{System Access}
VisIVOWeb is supported fully by commonly used Internet browsers, e.g. Internet Explorer, Firefox, Chrome, Safari and Opera. JavaScript,  Java, pop-up windows and cookies must be enabled.
\\ The system can be accessed using either of the following modes: authorized or anonymous. To employ an  authorized account, users are required to complete a web form; their system account is then activated following approval, typically in a couple of hours. On the other hand, anonymous accounts do not require approval, they are simply created through the home page of the system and they are activated immediately. These accounts  employ randomly generated usernames and empty passwords. Assuming there is sufficient amount of disk space available, there is no limit on the max number of anonymous accounts employed by the system.
\\All users (no matter if anonymous or authorized access) are provided with their own individual staging area within the system for uploading and managing not only their datasets but more importantly the visualizations produced from these datasets, e.g. images and movies. For anonymous accounts, although access to full functionality for processing and visualization is provided, they are valid for no more than four days since last user access. At the end of this period all user uploaded datasets and created visualizations are removed automatically. Authorized accounts are valid for two months since last user access. Prior to expiry a warning e-mail is sent, then a simple login into the system by the user can extend the validity of an account automatically. The typical scenario for importing datasets into the system is outlined in section 4.

\section{Importing Datasets}
Before any analysis and/or visualization is performed, user datasets must be imported into VisIVOWeb. The system currently supports most commonly employed astrophysical data formats e.g. ASCII and CSV tables, VOTables ~\cite{VOTABLE}, FITS Tables, GADGET-2 ~\cite{GADGET2}, Raw Binary and Raw Grid. Furthermore other data formats are supported either targeting specific applications that are used by a sizable scientific community or being of particular scientific interest. Consequently the system supports the outcomes of the code FLY ~\cite{FLY} and TVO XML, a VOTable that can be used to describe the content of generic binary files. This descriptor is currently under revision by the Theory Interest Group within the IVOA framework\footnote{http://www.ivoa.net/cgi-bin/twiki/bin/view/IVOA/IvoaTheory}.
\\Several interfaces are provided for importing depending upon the data format of the input datasets. VisIVOWeb can handle point clouds as well as volumetric datasets; the latter typically consist of collections of  scalar values arranged on regular grids. Furthermore point clouds can be transformed into volumetric datasets for visualization using a cloud in cell ~\cite{cloud-in-cell} smoothing algorithm.
\subsection{VisIVO Binary Tables}
Datasets imported into VisIVOWeb are always converted into an internal data representation called VisIVOBinaryTable (hereafter VBT). A VBT can describe both point datasets and volume datasets; for detailed information on its structure and header the reader is referred to ~\cite{VisIVOServer}. The importing of a dataset may produce one or more VBTs, e.g. a GADGET-2  snapshot can generate six VBTs corresponding to different particle types associated with gas, halo, disk, bulge, stars and boundary.
\subsection{Process}
The importing interface can be deployed by a local or a remote upload. The  local upload is  a synchronous operation to transfer a dataset from the user's PC to VisIVOWeb handled fully by the user's browser. However in case of large-scale datasets (e.g. more than 100 MBs) the suggested solution is a remote upload. This will perform an asynchronous transfer from a remote location to VisIVOWeb. The system supports most commonly used transfer protocols, e.g. http, sftp and ftp.
\\To import using remote upload, users are simply required to provide the address of a URL location in which the relevant datasets are stored. A  login and password can be provided optionally to grant access to the remote resources.  An e-mail is sent to the user at the end of the process reporting the details of the transfer operation. The overall time for importing consists of uploading, converting and creating the underlying database objects.
\begin {itemize}
\item  Uploading: This depends on the size of the data file to be processed and on the network bandwith.
\item Converting: This depends on the size of the data file and on the data type and number of tables that it contains. This operation is required for conversion of datasets into the internal data representation as described in section 4.1.
\begin{figure*}
\epsscale{1.4}
\plotone{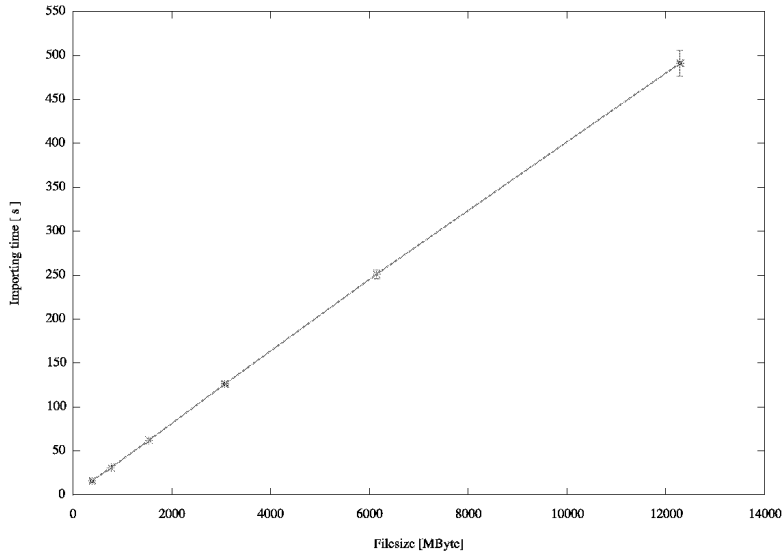}
\caption{Scaling of overall importing times against dataset sizes. The error bar is calculated from several repeated realizations.}\label{benchmark}
\end{figure*}
\item Creating Database Objects: This is typically a very short period of time that is required for creating all the necessary data structures and metadata related to importing. Section 5 describes the details of the database objects.
\end {itemize}
The graph in figure \ref{benchmark} shows the scaling of overall importing times versus dataset sizes. Our tests were performed on a dual core AMD Opteron processor 280 with 4GBs RAM. Several binary datasets were imported varying their sizes from 380 MBs up to 12 GBs. The linear scaling observed suggests that handling large datasets is effectively restricted only by underlying file system limitations.
\section{Data Management}
The underlying data-model is implemented through a relational database. This allows users their private staging area within the system for managing their datasets as well as images and movies produced from such datasets. Typically a number of operations are employed prior to any analysis and/or visualization, e.g. for extracting interesting features. This section describes the developed database objects (sections 5.1 and 5.2), the relevant interface (section 5.3) and outlines a representative number of available operations (section 5.4).
\subsection{Metadata Objects}
\begin{figure*}
\epsscale{2}
\plotone{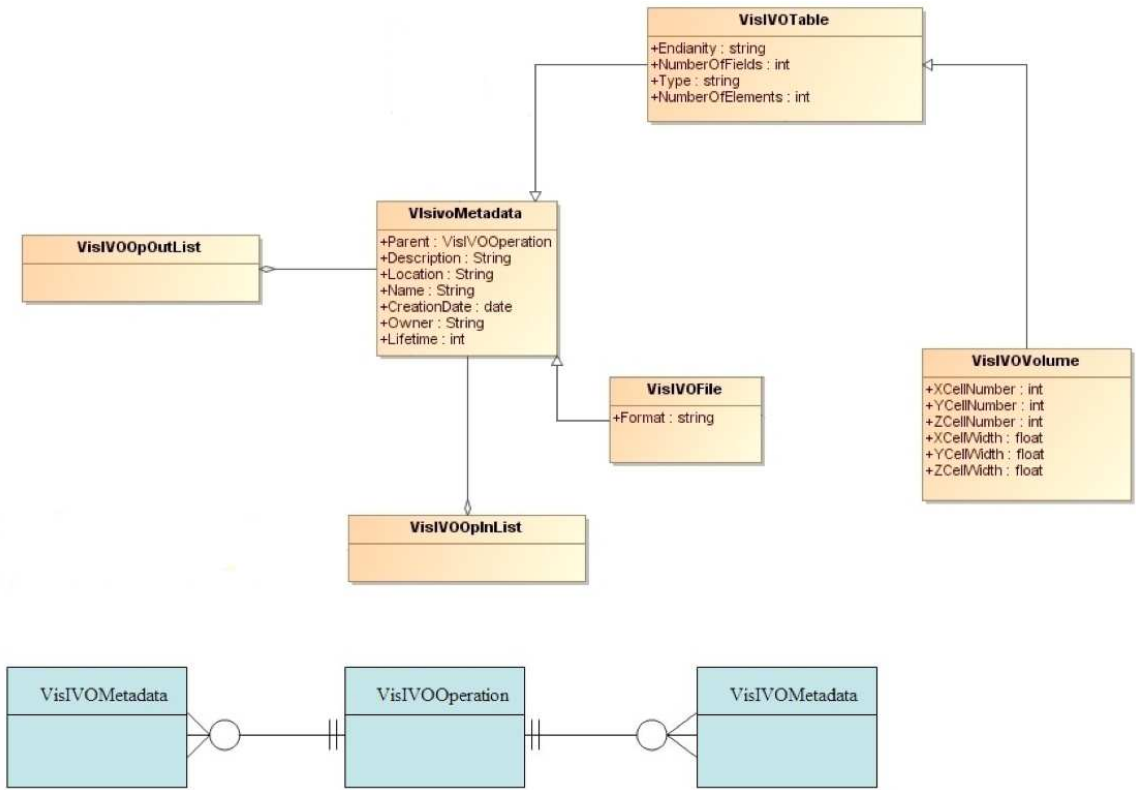}
\caption{The data model employed in VisIVOWeb: (a) UML class diagrams representing object hierarchy and (b) an entity relationship diagram (crow's foot notation). }\label{datamodel}
\end{figure*}

The metadata (or attributes) associated with uploaded datasets are encapsulated by customized VisIVOFile objects. These objects are inherited from VisIVOMetadata objects, that encapsulate attributes such as owner or creation time, while adding new attributes such as the file format employed (e.g. CSV or FITS Tables). Moreover the objects VisIVOTable and VisIVOVolume are defined by inheriting from VisIVOMetadata for representing VBTs (see section 4.1). VisIVOTable objects  introduce attributes such as endianity, number of fields, type and number of elements. VisIVOVolume objects introduce attributes such as number and size of cells for each dimension (Figure \ref{datamodel}). As an example consider importing into our system GADGET-2 datasets. A VisIVOFile object will be generated by the original file and importing may produce six (one for each particle type) VisIVOTable objects.
\subsection{Operation Objects}
An operation in VisIVOWeb can be described as a many-to-many relation connecting VisIVOMetadata objects. We can simplify this relation by introducing a VisIVOOperation object to treat a many-to-many relation as two one-to-many relations (Figure \ref{datamodel}). In other words many VisIVOMetadata objects can possibly generate several VisIVOMetadata objects by using a single VisIVOOperation. We also define the objects VisIVOOpOutList / VisIVOOpInList as aggregations of VisIVOMetadata objects. As an example consider merging VBTs representing point clouds. Then, VisIVOOpInList is the aggregation of the original VisIVOMetadata objects corresponding to the relevant VBTs while VisIVOOpOutList will contain a new VisIVOMetadata, i.e. the result of the operation. 
\subsection{Operational Scenario}
\begin{figure*}
\epsscale{0.7}
\plotone{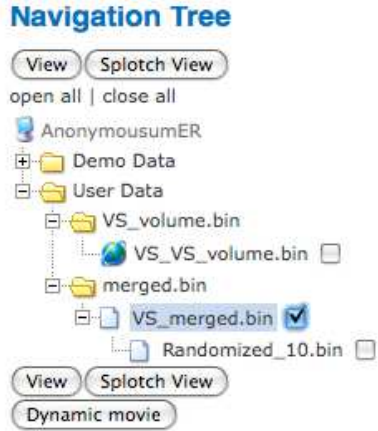}
\caption{The graphical interface employed in our system for data management activity.}\label{tree}
\end{figure*}

We provide our system with a graphical web interface  that allows to easily store data and  work with them. Data are organized in a tree-like structure (Fig. \ref{tree}). In this way a user is allowed to select one or more datasets and to use them as members of an operation or as elements of a visualization pipeline. Different objects in the data-model such as files, volumes and tables have different icons. This interface allows to manage object metadata and to select an individual  object or a whole branch to be deleted.\\
\subsection{Operations}
A single point VBT is required by the following operations:
\begin {itemize}
\item Randomization - A new point VBT is produced by randomly selecting particles according to user-prescribed percentage values.
\item Math - Several mathematical operations for combining columns of VBTs. A suitable drop-down menu is provided to assist users in inputting any required algebraic expression.
\item Select Rows - This is a multi-dimensional selector. The user can impose constraints to specify  min/max values for particular fields of a VBT. The result of the operation is a new point VBT matching the prescribed criteria.
\item Append - A new VBT is created from a number of VBTs with an identical number of columns.
\item Extraction - A new VBT is created by identifying a region of interest using a box/sphere. The user must supply suitable columns to be used as Cartesian coordinates, and also dimensions and placement of the extracting box/sphere. 
\item Point Distribution - A point VBT is transformed into a volumetric VBT. Each point is distributed to nearby voxels using one of the following algorithms: cloud in cell, nearest grid point or triangular shape cloud ~\cite{cloud-in-cell}.
\end {itemize}
A point VBT and a volume VBT are required by the following operation:
\begin {itemize}
\item Grid2Point - For distributing the values of a volumetric VBT within a point VBT in the same domain using the cloud in cell, nearest grid point or triangular shape cloud algorithm.   
\end {itemize}    
The following operations are for volume VBTs only:
\begin {itemize}
\item Coarse Volumes - This reduces a VBT using a plane for selecting a region of interest according to user-prescribed percentage values.
\item Sub Volumes - This operation extracts rectangular regions of interest from prescribed VBTs.
\end {itemize}
Currently several new operations are planned to be implemented; to obtain the most recent developments the reader is referred to  Sourceforge.net\footnote{\url{http://sourceforge.net/projects/visivoweb/}}.
\begin{table*}[ht]
\caption{Modulating VisiVOWeb rendered views for point clouds.}
\begin{tabular}{c c c c c c c c}
&  & & & & & &\\
\hline
& X  & Y & Z & Color & Intensity & Radius & Height\\
\hline
VTK-Based & * & * & * & * &  &* &*\\
Splotch & * & * & * & * & * &&\\
\hline
\end{tabular}
\label{PointDataFeatures}
\end{table*} 
\section{Visualization}
The standard visualization views in VisIVOWeb employ orthographic projections.  The focal point is fixed in the center of the box to be observed and users can prescribe the camera azimuth and elevation. The system supports several algorithms based on the Visualization Toolkit\footnote{\url{http://www.vtk.org/ }} and Splotch ~\cite{Splotch1, Splotch}. The former is suitable for point clouds and volumetric datasets while the latter is a customized ray-tracer suitable for particle-based datasets.
To decouple our visualization engine from the hosting  computer infrastructure a default installation of VisIVOWeb employs the Mesa libraries\footnote{\url{http://www.mesa3d.org/ }}, an open-source implementation of OpenGL. Nevertheless, a customized installation of the system is feasible so as to exploit fully the hardware capability of the underlying hosting server.
\subsection{Point Clouds}
\begin{figure*}
\epsscale{1.2}
\plotone{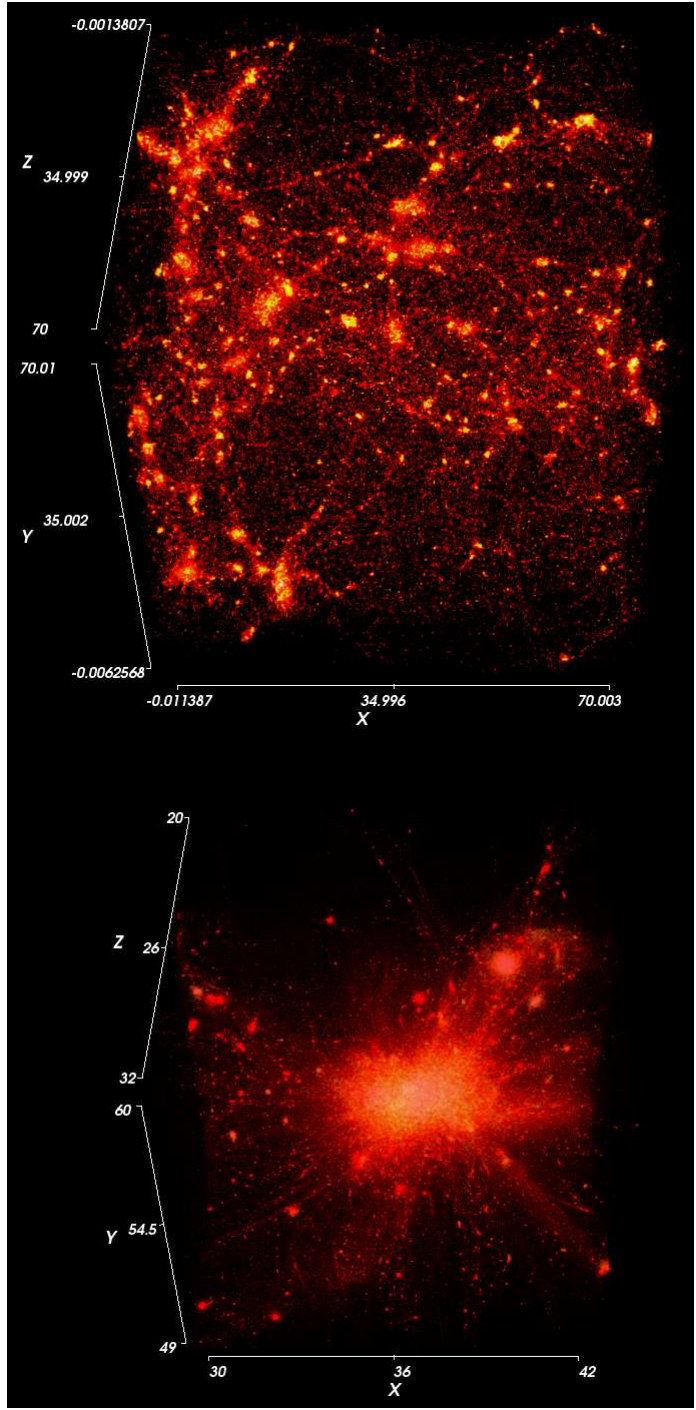}
\caption{The upper image is the entire computational box of an N-body simulation z = $0.1$, box size = $70 Mpc/h$, number of  particles = $800^3$. The lower image is a sub-box of the same simulation. The electronic version of the article includes color plates.}\label{VTKExample}
\end{figure*}

Point clouds can be visualized using either VTK-based algorithms or Splotch, provided that suitable columns are selected from the input VBT for a Cartesian system of coordinates to be used for rendering. The entire computational box of an N-body simulation (z = 0.1) is displayed at the top of figure \ref{VTKExample}; its size is 70 Mpc/h. The image at the bottom illustrates fine details in a sub-box from the same simulation. By using predefined color tables, scalar values can be mapped to colors so as to modulate VisIVOWeb rendered views based on values of specific properties of individual points in a cloud. Apart from colors, user defined shapes, e.g. cylinders, cones or spheres, can also be attached to particular points so that geometrical characteristics, e.g. height or radius, can be modulated according to values of specific point properties. Splotch also operates by mapping scalar data (a column of the input VBT) to colors by indexing though a set of predefined color palettes. A summary of mappings of VBT fields to visible elements in the VisIVOWeb views is shown in table \ref{PointDataFeatures}, for VTK-based and Splotch rendering respectively.
\begin{figure*}
\epsscale{1.4}
\plotone{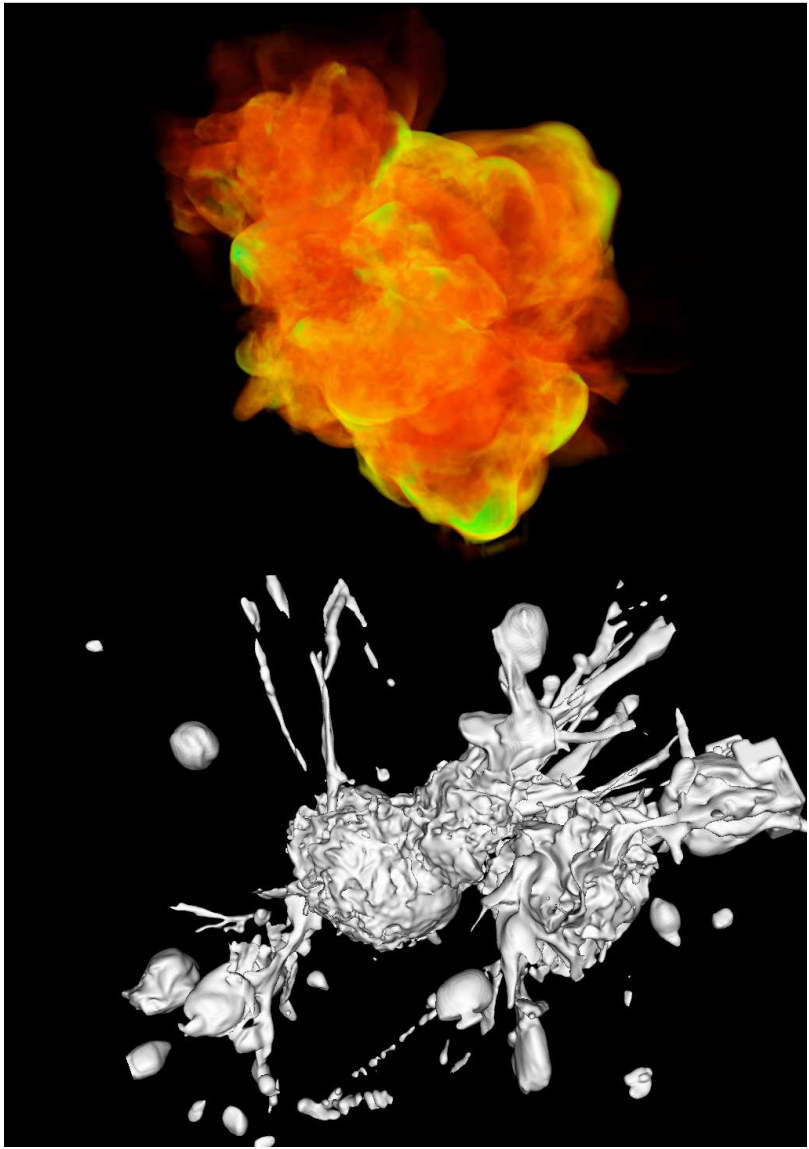}
\caption{A major merger galaxy cluster simulated with high spatial resolution using the ENZO 1.5 cosmological code. The upper image is a temperature profile visualized using volume rendering. The
lower image is a density profile visualized using isosurfacing. Box size = $187Mpc/h$; voxels = $544^3$.The electronic version of the article includes color plates.}\label{Volume}
\end{figure*}
\begin{figure*}
\epsscale{1.3}
\plotone{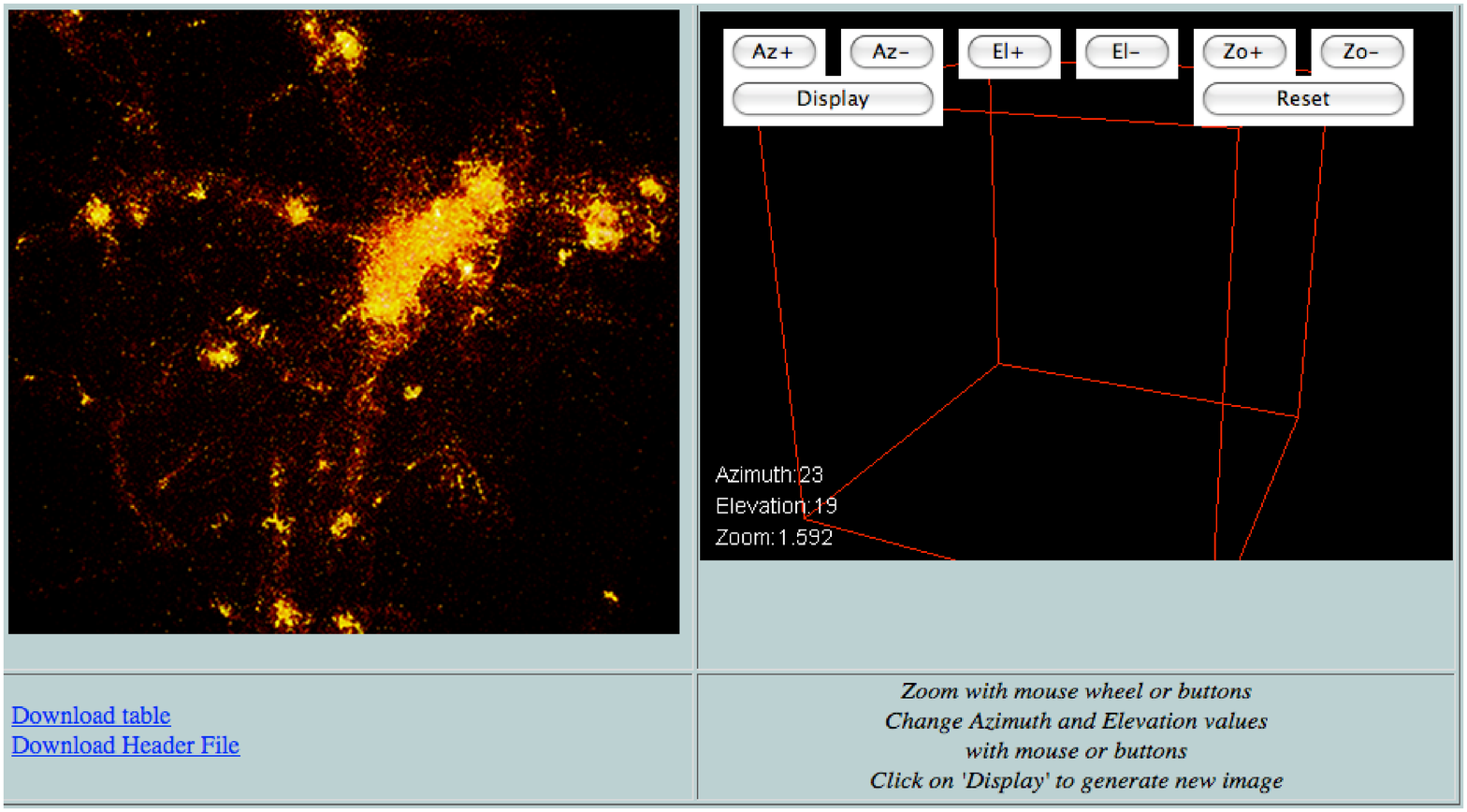}
\caption{The interactive widget employed for adjusting rendering settings for a cubical sub-sample in an N-body cosmological simulation (z = 0.1).}\label{MagicBox}
\end{figure*}
\subsection{Volumes}
Volumetric datasets can be visualized by using either volume rendering algorithms or isosurfacing.  Figure \ref{Volume} illustrates the renderind of a major merger galaxy cluster simulated with high spatial resolution using the ENZO 1.5\footnote{http://lca.ucsd.edu/portal/software/enzo} cosmological code. The upper image is a temperature profile visualized using volume rendering. The lower image is a density profile visualized using isosurfacing. (Box size = $187Mpc/h$; voxels = $544^3$ ). The isosurfaces in the bottom image correspond to prescribed constant values attained by the density function. Inspecting the rendering results for a number of different constant values reveals the inner structure complexity of the underlying datasets.
\begin{figure*}
\epsscale{1.2}
\plotone{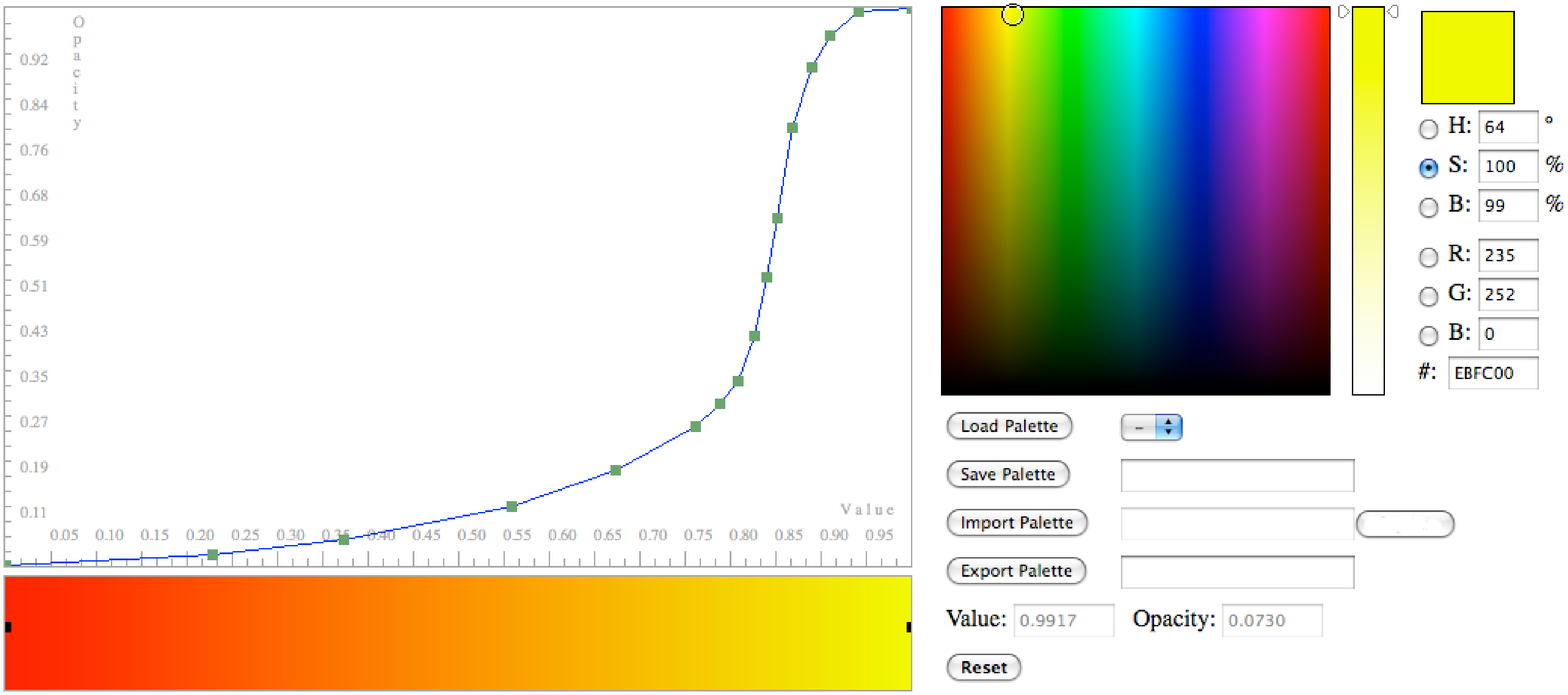}
\caption{The widget employed for adjusting opacity functions and color maps.}\label{ColorMapEditor}
\end{figure*}
\section{Visualization Interfaces}
The visualization interfaces provided within VisIVOWeb are controlled by a widget interactively for setting viewing parameters, e.g. zoom in/out factors, azimuth and elevation (Figure \ref{MagicBox}). This widget is implemented as a Java applet controlling an interactive box within the user's browser. The user can interact with the widget by using either the computer mouse or the buttons built into the widget. Once the viewing parameters are fixed the user can render via the display button.

\subsection{Opacity Function and Color Maps}
An interactive editor is provided (Figure \ref{ColorMapEditor}) allowing users to create customized color maps and opacity functions for visualization. On the left hand side of the interface the opacity function and color map are displayed. Using the computer mouse a user can edit this opacity function not only by inserting additional nodes but also by updating or even deleting existing nodes, simply by dragging them out of the window. The overall opacity function is constructed by linearly interpolating all nodes. The user can also adjust each node's color through the color selector, HSB editor or RGB editor, the corresponding color map is updated in real time. Color maps and opacity functions can be saved for later use into system templates.

\subsection{Interfaces}
\begin{figure*}
\epsscale{1}
\plotone{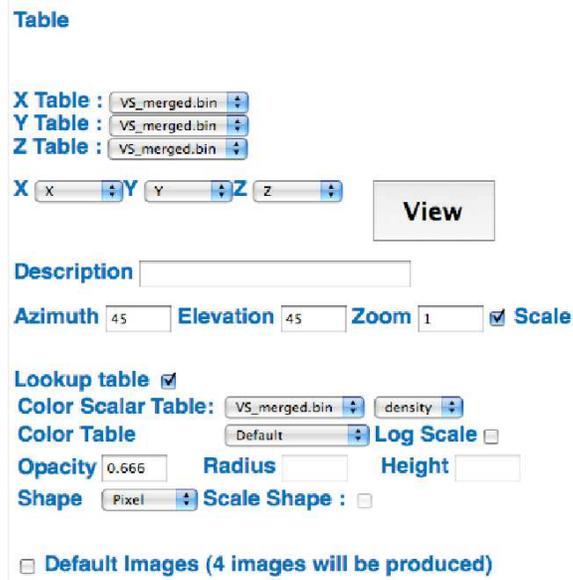}
\caption{The graphical user interface employed in VisIVOWeb for VTK-based rendering.}\label{VTKVis}
\end{figure*}

\begin{figure*}
\epsscale{1.2}
\plotone{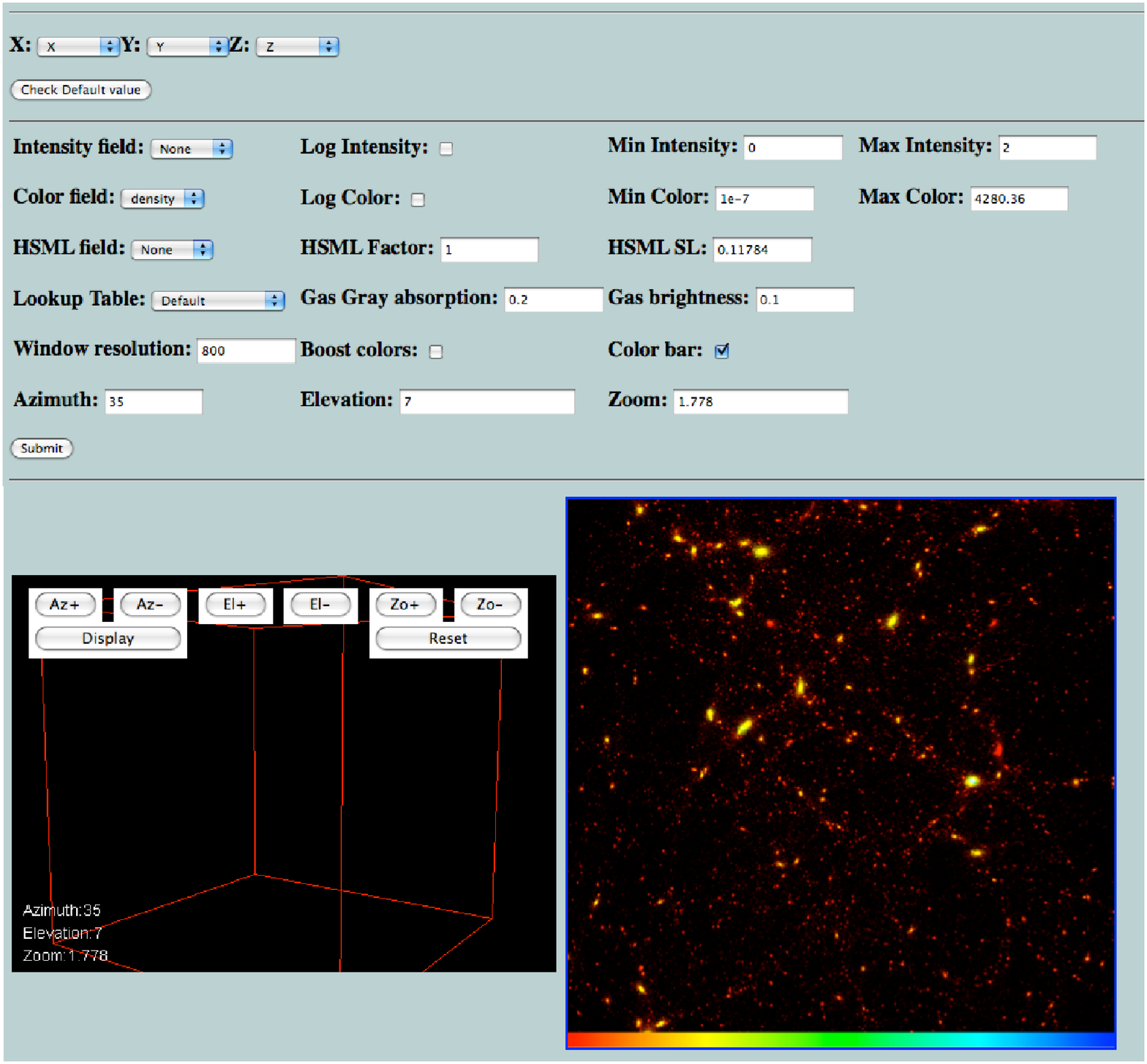}
\caption{The graphical user interface employed in VisIVOWeb for Splotch rendering.}
\label{SplotchVis}
\end{figure*}

The graphical user interface for VTK-based rendering is shown in Figure \ref{VTKVis}. The user can select VBT columns to construct a Cartesian system of coordinates for rendering. The scale option is employed for transforming coordinates so that rendering occurs within a cubical box if required. Changes in the opacity value are typically used for revealing and/or masking a dataset's inner structures. Finally, using the lookup table option a VBT column can be selected to be mapped into a color palette. A set of pre-defined color tables are accessible by default. The user can also apply shapes, e.g. spheres, cylinders or cones, during rendering of individual points, modulating their geometric characteristics by employing a specific property.
The graphical user interface for Splotch rendering is illustrated in Figure \ref{SplotchVis}. Firstly, the VBT columns for constructing the Cartesian system of coordinates for rendering must be chosen. Subsequently several scalar values need to be selected by the user to be associated to intensity and color, HSML (SPH SMoothing Length) ~\cite{SPH}, HSML factor, gas gray absorption and brightness respectively.
\section{Scientific Movies}
Scientific movies are useful not only to scientists for presenting and communicating their research results, but also to museums and science centers for introducing the general public into complex scientific concepts. The creation of a movie represents a significant challenge for the underlying computational resources as often hundreds or thousands of high quality images must be produced. For example, creating a fly-through lasting 5 mins within a 16 million particle cosmological simulation requires 38200 seconds in a dual core AMD Opteron processor 280 with 4GBs of RAM memory.
\subsection{Temporal Parallelism }
To alleviate the high computational demands a temporal parallelism is adopted for decomposing the overall computation. The fundamental unit of computation is a complete frame and an individual processor is assigned with a number of frames to be rendered. Our system is fully integrated with  OpenPBS\footnote{\url{http://www.platform.com/ }} and LSF\footnote{\url{http://www.openpbs.org/ }}. A number of frames from scientific movies created on a cluster of 54 CPU cores using the portal at  ~\cite{INAF-Catania} are shown in figure \ref{Volume}.
\subsection{Operational Scenario}Currently movies are created in VisIVOWeb by using several interfaces. Firstly, users can create fly-throughs by moving a camera along a motion path which is prescribed within the domain of a dataset.\\
Secondly, fly-throughs can be produced by intermediate snapshots specified as camera positions / orientations; the system generates a movie with a camera path containing the specified positions / orientations.\\
Thirdly, movies can be created by interpolating several steps of a time evolution. The user can browse a cosmological time evolution and choose two or more coherent datasets. The  system will then produce the necessary number of intermediate datasets by calculating particle positions and applying boundary conditions as necessary.  This approach can be very useful, e.g. in revealing galaxy formation or large-scale structures such as galaxy clusters.

\subsection{A Movie from a 512 Million Particle Simulation}
One of the most challenging tasks in performing large-scale cosmological simulations is monitoring the outcomes, so that if anomalies are observed the simulation run is corrected appropriately. Typically such monitoring requires visualization of several snapshots at different red-shift, each being tens of GBs.\\
We recently performed a large-scale structure of the Universe simulation (100Mpc $h^{-1}$) containing $800^{3}$ particles.
The run was performed on the Cometa Consortium Grid\footnote{\url{http://www.pi2s2.it}} using FLY ~\cite{FLY}. The simulation was carried out over three months using 250 CPU cores; each data snapshot was 12 GBs.\\
We employed VisIVOWeb firstly to extract 8 million particles and secondly to identify a sub-region (15Mpc $h^{-1}$) by using the operations described in section 5.4.  We then interpolated the original snapshots in the time domain producing  a large number of intermediate frames that have been used in creating a movie showing the fluid time evolution. Overall 1500 intermediate datasets were created requiring a CPU time of 2 months. The final movie employs a frame rate of 10 fps and can be found at \url{http://astrct.oact.inaf.it/visivo/cosmo/}.  The whole data archive handled for the movie creation was in excess of 18 TBs.  
\section{VisIVOWeb Connect and Network}
VisIVOWeb Connect is a lightweight solution that allows seamless connection of VisIVOWeb to any astrophysical archive by adopting an http based approach. Using this technology users can view and manage datasets from any Internet archive directly within VisIVOWeb. The system can handle any data archive and it allows automatic uploading of datasets into a user-specified VisIVOWeb portal.\\
To exploit the VisIVOWeb Connect functionality the relevant datasets must be encoded using a supported data format and be accessible through a supported data transfer protocol. Assuming that these conditions are observed, integrating VisIVOWeb Connect with any Internet archive is straightforward, simply requiring a URL containing locations / datatypes of the datasets to be uploaded.\\
Recent developments begun the deployment of a network of VisIVOWeb portals. At the time of writing active nodes exist at ~\cite{INAF-Catania}, ~\cite{INAF-Trieste} and ~\cite{Portsmouth}. The creation of a network of VisIVOWeb portals allows the user to choose the closest in order to minimize uploading times. Furthermore it allows to potentially customize different portals with tools specific to particular archives. 

\section{Current Status and Future Work}
The visualisation functionality supported by VisIVOWeb is rich and we are currently in the process of developing several demonstrators to be used for formal evaluation in the context of specific astrophysical communities using numerical simulations\footnote{\url{http://astrct.oact.inaf.it/visivo/cosmo/}} but also real-world observations\footnote{ \url{http://www.sdss.org.uk/}}. The initial feedback is very encouraging and we are working on fixing software bugs prior to our next major release, but also pursuing several enhancements.\\
Firstly, several new operations are planned to be implemented, to expand significantly the collection of operations presented in section 5.4; the reader is referred to sourceforge.net \footnote{\url{http://sourceforge.net/projects/visivoweb/}} for the newest developments. Secondly, we are working in upgrading the core visualisation functionality to the latest version of the Visualisation Toolkit, and the recently released version of Splotch ~\cite{Splotch}. The latter work is of particular importance as it offers the exciting possibility for exploiting hybrid massively-parallel architectures containing large numbers of multicore CPUs and CUDA \footnote{ \url{http://www.nvidia.com/object/cuda_home_new.html}} enabled GPUs. To exploit high performance computing, we have already integrated within our system the functionality to allow authenticated users to submit movie creation jobs using the EGEE grid resources, based on the computational capability provided by the grid operated by the COMETA consortium.\\
Finally, depending upon the user feedback obtained during the formal evaluation activity and within the scope of specific astrophysical communities, we are planning to incorporate new functionality, e.g. for rendering streamlines and AMR datasets efficiently or designing highly customized opacity functions. Such algorithms either already exist in the underlying visualization libraries (but they are simply not enabled) or they can be developed readily. This work will also dictate the direction of development for customizing future VisIVOWeb portals employing VisIVOWeb Connect.  At the time of writing, this technology is used for connecting our system seamlessly with the Italian Theoretical Virtual Observatory web archive  ~\cite{ITVO} and the UK mirror of the Sloan Digital Sky Server (data release 7) CasJobs service ~\cite{SLOAN-UK}.\\
There is a number of user guides for VisIVOWeb available to download from the nodes operating currently ~\cite{INAF-Catania,INAF-Trieste, Portsmouth}. Each node also contains streaming videos of on-line tutorials, the relevant slides are available for download at the VisIVO portal\footnote{\url{http://visivo.oact.inaf.it/}}. The underlying visualization technology documents are accessible at the svn repository. The functionality of our system has been presented in the VO-Day\footnote{\url{http://wwwas.oats.inaf.it/voday/}} events run recently accross Italy, organized in the framework of Euro VO-AIDA\footnote{\url{http://cds.u-strasbg.fr/twikiAIDA/bin/view/EuroVOAIDA/WebHome/}} and  IVOA.

\section{Summary}
Modern visualization can aid astrophysicists in gaining good insights of complex datasets, generated either from real-world observations or numerical simulations, by providing suitable tools for rapid and intuitive visual discovery. We have introduced an open-source WWW environment called VisIVOWeb for visualization of large-scale astrophysical datasets offering astrophysicists an array of advanced visualization tools that can be exploited effectively using common Internet browsers. The system offers full data management functionality so that users can upload and manage their datasets, e.g. by using interactive widgets to construct customised renderings or generating meaningful astrophysical animations or even storing their results (movies or otherwise) for future reference. To the best of our knowledge, there are no other comparable open-source environments available to astrophysicists in the Virtual Observatory.\\
We reviewed relevant work and introduced the way users can gain access to the system for importing their datasets. We discussed the underlying model for data management, focusing on the developed database objects, the relevant interface and a number of currently available operations. We also presented the visualization algorithms supported by the system, founded on the Visualization Toolkit and Splotch, together with the relevant visualization interfaces including an interactive editor for constructing customized opacity functions. The operational scenario for generating scientific movies from large-scale datasets was then introduced discussing a specific example for demonstration based on a large-scale simulation containing 512 million particles. The whole data handled by VisIVOWeb for this movie creation was in excess of 18 TBs. We then introduced VisIVOWeb Connect/Network for networking VisIVOWeb portals and connecting them seamlessly to existing astrophysical archives. Finally we concluded with a description of the current status and pointers to a number of future developments.

\acknowledgments
The authors would like to thank Marco Comparato for his contribution in realizing the first version of VisIVOWeb. We would also like to thank Patrizia Manzato and Marco Molinaro for their intensive debugging activity during initial development of VisIVOWeb.\\
Original datasets in figure \ref{Volume} were provided by the IRA-CINECA Simulated Cluster Archive\footnote{ \url{http://data.cineca.it/}}. This work has made use of results produced by the PI2S2 project\footnote{url{http://www.pi2s2.it}} managed by the Consorzio COMETA\footnote{\url{http://www.consorziocometa.it}}, co-funded by the Italian Ministero dellÕIstruzione, UniversitaÕ e Ricerca (MIUR) under Piano Operativo Nazionale Ricerca Scientifica, Sviluppo Tecnologico, Alta Formazione (PON 2000-2006).
Finally we thank Luigia Santagati for proof-reading.

\end{document}